\documentclass[aoas,nameyear,seceqn,dvips]{arximspdf}
\usepackage{graphics}
\doi{10.1214/09-AOAS258}
\volume{3}
\issue{4}
\pubyear{2009}
\firstpage{1542}
\lastpage{1565}

\makeatletter
\def\epsilon{\varepsilon}

 \newtheorem{prop}{Proposition}

  \let\sv@tabnotetext\tabnotetext
  \let\sv@tabnotemark@fmt\tabnotemark@fmt
   \long\def\legend#1{{\let\tabnote@indent\leavevmode\sv@tabnotetext[]{}{#1}}}

\makeatother

\begin{document}
\begin{frontmatter}

\title{Profiling time course expression of virus genes---an
illustration of Bayesian inference under shape restrictions\protect\thanksref{TITLE1}}
\runtitle{Genome-wide expression profile of virus genes}
\thankstext{TITLE1}{Supported by Taiwan NHRI Grant
BS-095-PP-05 and Taiwan NSC Grants NSC 94-3112-B-400-002-Y (ISC)
and NSC 95-3112-B-400-005-Y (CAH).}

\begin{aug}
\author[A]{\fnms{Li-Chu} \snm{Chien}\thanksref{aut1}},
\author[B]{\fnms{I-Shou} \snm{Chang}\thanksref{aut1}},
\author[C]{\fnms{Shih Sheng} \snm{Jiang}\thanksref{aut1}},
\author[A]{\fnms{Pramod K.} \snm{Gupta}},
\author[D]{\fnms{Chi-Chung} \snm{Wen}},
\author[E]{\fnms{Yuh-Jenn} \snm{Wu}}
and
\author[A]{\fnms{Chao A.} \snm{Hsiung}\corref{}\ead[label=e7]{hsiung@nhri.org.tw}}
\runauthor{L.-C. Chien et al.}
\thankstext{aut1}{These authors contribute equally and are joint first authors.}
\affiliation{National Health Research Institutes,
National Health Research Institutes, National Health Research Institutes, National Health Research Institutes, Tamkang University,
Chung Yuan Christian University and National Health Research Institutes}
\address[A]{L.-C. Chien\\
P. K. Gupta\\
C. A. Hsiung\\
Division of Biostatistics\\
\quad and Bioinformatics\\
National Health Research Institutes\\
35 Keyan Road, Zhunan Town,\\
\quad Miaoli County 350\\
Taiwan\\
\printead{e7}} 
\address[B]{I-S. Chang\\
Division of Biostatistics\\
\quad and Bioinformatics\\
and\\
National Institute of Cancer Research\\
National Health Research
Institutes\\
35 Keyan Road, Zhunan Town,\\
\quad Miaoli County 350\\
Taiwan}
\address[C]{S. S. Jiang\\
National Institute of Cancer Research\\
National Health Research
Institutes\\
35 Keyan Road, Zhunan Town,\\
\quad Miaoli County 350\\
Taiwan}
\address[D]{C.-C. Wen\\
Department of  Mathematics\\
Tamkang University\\
151 Yingchuan Road, Tamsui Town,\hspace*{20pt}\mbox{}\\
\quad Taipei County 251\\
Taiwan}
\address[E]{Y.-J. Wu\\
Department of Applied Mathematics\\
Chung Yuan Christian University\\
200 Chung Pei Road, Chungli City 320\\
Taiwan}
\end{aug}

\received{\smonth{2} \syear{2008}}
\revised{\smonth{12} \syear{2008}}

\begin{abstract}
There have been several studies
of the genome-wide temporal transcriptional program of viruses, based
on microarray experiments, which are generally useful in the
construction of gene regulation network. It seems that biological
interpretations in these studies are directly based on the
normalized data and some crude statistics, which provide rough
estimates of limited features of the profile and may incur biases.
This paper introduces a hierarchical Bayesian shape restricted
regression method for making inference on the time course
expression of virus genes. Estimates of many salient features of
the expression profile like onset time, inflection point, maximum
value, time to maximum value, area under curve, etc. can be
obtained immediately by this method. Applying this method to a
baculovirus microarray time course expression data set, we
indicate that many biological questions can be formulated
quantitatively and we are able to offer insights into the baculovirus
biology.
\end{abstract}


\end{frontmatter}

\section{Introduction}\label{sec1}

\subsection{Transcription program of virus}\label{sec1.1}

With a custom made baculovirus DNA microarray, Jiang et al. (\citeyear{Jiaetal2006})
investigated the temporal transcription program of one of the best
characterized baculoviruses, Ac\textit{M}NPV, in its host
lepidopteran Sf21 cells. They uncovered sequential viral gene
expression patterns, which are possibly regulated by different
mechanisms during different phases of infection, compared the
transcription profile of a mutant virus with that of the wild
type, and suggested that the array strategy taken in the study
points to a very productive direction for constructing a
baculovirus gene regulation network.

The experiments of Jiang et al. (\citeyear{Jiaetal2006}) are briefly summarized as
follows. They use single color cDNA microarray experiments with
external controls for data normalization. Each chip has exactly
four spots for each of the 156 open reading frames, referred to as
genes henceforth, of baculovirus; total RNA samples of baculovirus
genes were taken at several different time points during the 72
hours following infection; the sample for each time point is
hybridized to a single chip. The normalized time course expression
data are shown to be in good agreement with those obtained by
the real-time PCR method for five randomly chosen genes; the data for
each gene used in the study of temporal transcription is based
solely on the normalized expression levels at these time points
and on the crude estimates of its onset time and the time that its
expression attains its maximum.

A rough idea regarding virus gene expression is that genes of a
virus have their time course expression level being zero
initially, then increasing after a while and finally decreasing;
because viruses do not have their own machinery for gene
transcription, their genes start to express only after getting
into cells, and cells may eventually malfunction when infected. It
is of interest and feasible to make use of this idea to profile
the time course expression of each virus gene, based on microarray
data, to estimate salient features of the profile like onset time,
time to maximum value, maximum value, area under the profile
curve, etc. and to test the shape hypotheses on the profile curve
like unimodality on certain time intervals.  It is hoped that this
approach to gene expression analysis of viruses would eventually
provide a sound basis for the study of the temporal transcription
program of viruses.

The purpose of this paper is to propose a Bayesian shape
restricted regression model based on the above property of a
virus, illustrate this model by profiling the time course
expression of genes of baculovirus, and indicate that this
approach does provide more insights into baculovirus, compared
with the crude statistics used in Jiang et al. (\citeyear{Jiaetal2006}). Among
others, a prominent example in this regard is that this new
approach seems to support the widely accepted conjecture that
structural genes of the virus may have a larger amount of total
expression level, which is hard to examine by the method in Jiang
et al. (\citeyear{Jiaetal2006}).

This method is illustrated on the dataset for the baculovirus
Bac-PH-EGFP in Jiang et al. (\citeyear{Jiaetal2006}). With 16 time points, this
dataset seems to hold a promising opportunity to capture the main
features of the transcription profile.  We note that the other two
datasets in Jiang et al. (\citeyear{Jiaetal2006}) have only 6 time points and 5 of
them are in the initial two hours post infection and it is hard to
infer some of the main features of the profile based on them.

Because microarray experiments offer feasible approaches to the
studies of the genome-wide temporal transcriptional program of
viruses, which are generally useful in the construction of gene
regulation network, there have been many genome-wide expression
studies of virus genes. See, for example, Yang et al. (\citeyear{Yanetal2002}),
Iwanaga et al. (\citeyear{Iwaetal2004}), Duplessis et al. (\citeyear{Dupetal2005}), van Munster et al.
(\citeyear{Munetal2006}), Majtan et al. (\citeyear{Majetal2007}), Smith (\citeyear{Smi2007}) and references therein;
they considered different viruses and/or different host cells. It
seems that all the biological interpretations in these studies are
directly based on the normalized data and crude statistics, which
seem to provide only naive estimates of limited features of the
profile, and there are some discrepancies reported in the
literature; see, for example, Smith (\citeyear{Smi2007}). It is of great
interests to compare the transcriptional studies based on
different but related strains of viruses and/or different and
related host cells so as to build a gene regulation network. We
note that comprehensive comparisons depend on comprehensive and
rigorous time course expression profiling of genes in each study.
The focus of this paper is the latter.

\subsection{Statistical modeling strategy}

Preliminary examination of the Bac-PH-EGFP data suggests that two
of the 156 genes seem to have their expression levels being zero
finally as well as initially and the rest of the 154 genes being zero
only initially, probably because no data were taken at time point
beyond 72 hours and the life cycle of baculovirus is longer than
72 hours, according to Friesen and Miller (\citeyear{FriMil2001}). To make the
presentation concise, we limit our attention to these 154 genes in
this paper; the other two genes can be studied similarly.

Let $\mathcal A$ denote the set of all smooth functions on $[0,1]$
that are zero initially, start to increase after a while, and stay
positive onward. The task of profiling the time course expression
level of virus genes will be considered a shape restricted
regression problem with the regression function belonging to
$\mathcal A$. Let $g=1,2,\ldots,154$ index the 154 genes of the
baculovirus. For $g=1,\ldots,154$, we assume that, given $F_g$ in
$\mathcal A$,
\begin{equation}\label{eq1.1}
Y_{jkg}=F_{g}(X_k)+\epsilon_{jkg}.
\end{equation}
Here $\{X_{k}\mid k=0,\ldots,K\}$ are constant design points in
$[0,1]$,
$\{Y_{jkg}\mid j=1,\ldots,m_k,k=0,\ldots,K,g=1,\ldots,154\}$ are
response variables, and for every
$j=1,\ldots,m_k,k=0,\ldots,K,g=1,\ldots,154$, $\epsilon_{jkg}$ are
independent normal errors with mean $\mu_{g}$ and variance
\begin{equation}\label{eq1.2}
\sigma_{kg}^{2} =\sigma_{g}^2\bigl( F_{g}(X_{k})+\mu_{g}\bigr)^{\xi_{g}}
\end{equation}
for some $\xi_{g}=0$, 1 or 2.

In this paper $X_{k}$ represents a time point at which the mRNA
sample is taken for microarray experiments; $Y_{jkg}$ is the
expression level, in terms of fluorescent intensity, obtained at
the $j$th spot of the $g$th gene for the sample taken at time
point~$X_k$. More specifically, in our data, let $[0,1]$ denote the
time period of 72 hours, then $K=15$, $m_k=4$,
$(X_0,X_1,\ldots,X_{15})=(0, 1/216, 1/108, 1/72, 1/36, 1/24,
1/12, 1/8, 1/6, 5/24, 1/4, 1/3, 5/12, 2/3, 5/6,\break 1)$.

The variance structure in (\ref{eq1.2}) is a simple way to take into
consideration the observation that for single color cDNA
microarray experiments, larger intensities often incur larger
variances when considering replicates. The reason for not assuming
$\epsilon_{jkg}$ having zero mean is that there are always
background intensities due to nonspecific hybridization and, hence,
$E(Y_{jkg})$ may not be zero even when the expression level
$F_g(X_k)$ is zero.

We now explain that Bernstein polynomials can be used to study the
above shape restricted regression model. For integers $0 \leq i
\leq n$, let $\varphi_{i,n}(t)=C^n_i t^i(1-t)^{n-i}$, where
$C^n_i=n!/(i!(n-i)!)$. The set
$\{\varphi_{i,n}\mid i=0,\ldots,n\}$ is called the Bernstein basis
for polynomials of order up to $n$. Let $\mathcal{B}=[0,1]\times
\bigcup^{\infty}_{n=3} (\{n\}\times \mathbb{R}^{n-1})$. Define
$\mathbf{F}\dvtx\mathcal{B} \times [0,1] \longrightarrow \mathbb{R}^1 $ by
\begin{equation}\label{eq1.3}
\mathbf{F}(c, n, b_{2,n},\ldots,b_{n,n};t)=\sum^{n}_{i=2}b_{i,n}\varphi_{i,n}\biggl(\frac{t-c}{1-c}\biggr)I_{(c,1]}(t),
\end{equation}
where $(c,n, b_{2,n},\ldots,b_{n,n})\in {\mathcal{B}}$
and $t\in [0,1]$. We also denote (\ref{eq1.3}) by ${F}_{c,b_{n}}(t)$ if
$b_{n}=(b_{2,n},\ldots,b_{n,n})$. We will see in Section \ref{sec2} that
${F}_{c, b_{n}}(\cdot)$ is a member of $\mathcal A$ if $0\leq
\min_{l=2,\ldots,n} b_{l,n}< \max_{l=2,\ldots,n}
b_{l,n},$ and every member of $\mathcal A$ can be approximated by
${F}_{c, b_{n}}(\cdot)$ satisfying these restrictions on $b_{n}$.
This observation suggests that, by means of (\ref{eq1.3}), Bernstein
polynomials form a useful tool to introduce priors on $\mathcal A$
for a Bayesian analysis.

We will consider Bayesian hierarchical models based on (\ref{eq1.3}). With
priors on a space of smooth functions satisfying certain shape
restrictions and parameters in the priors based on crude estimates
from data, our approach has the advantage of utilizing prior
knowledge from biology; with 154 correlated and possibly similar
profiles to study, hierarchical regression models take advantage
of the possibility of data driven shrinkage-type estimates.

We note that Bayesian shape restricted inference with priors
introduced by Bernstein polynomials was studied by Chang et al.
(\citeyear{Chaetal2005}), which provides a smooth estimate of an increasing failure
rate based on right censored data, and by Chang et al. (\citeyear{Chaetal2007}),
which compares the Bernstein polynomial method with the
density-regression method [Dette, Neumeyer  and Pilz (\citeyear{DetNeuPil2006})] in estimating an
isotonic regression function and a convex regression function. It
was also shown there that these Bayesian estimates perform
favorably, in addition to the facts that these priors easily take
into consideration geometric information, select only smooth
functions, can have large support, and can be easily specified. We
note that Petrone (\citeyear{Pet1999}) made use of these nice properties in her
study of random Bernstein polynomials and for sampling the
posterior distribution, proposed algorithms that regards the
construction of the Bernstein--Dirichlet prior as a histogram
smoothing.

The present paper indicates that the expression profiles of virus
genes can also be efficiently studied by random Bernstein
polynomials, making use of the shape restrictions described above.
We will estimate salient features of the profile like onset time,
inflection point, maximum value, time to maximum value, area under
the profile, etc., utilizing the fact that the derivative of a
polynomial has a closed form. We will also test the hypothesis on
the shape of the time course expression profile; for example, we
will examine whether it is unimodal on the region $[0,\tau]$ for
some $\tau<1$. In fact, by calculating both the posterior
probability and the prior probability that it is unimodal on
$[0,\tau]$, we offer an assessment of the strength of the evidence
in favor of the hypothesis. We note that this direct approach to
hypothesis testing is markedly different from the frequentist
$p$-value approach, as discussed in Kass and Raftery (\citeyear{KasRaf1995}) and
Lavine and Schervish (\citeyear{LavSch1999}), for example.

There is a large literature on shape restricted inference since
Hildreth (\citeyear{Hil1954}) and Brunk (\citeyear{Bru1955}). Most of them treat isotonic and
concave regressions from the frequentist viewpoint. Readers are
referred to Gijbels (\citeyear{Gij2003}) for an excellent review and to Dette, Neumeyer  and Pilz
(\citeyear{DetNeuPil2006}) for some of the more recent developments. For the Bayesian
approach, there are the works of Lavine and Mockus (\citeyear{LavMoc1995}), Dunson
(\citeyear{Dun2005}) and Chang et al. (\citeyear{Chaetal2007}), among others. This paper
illustrates the use of the Bernstein polynomial in investigating the
strength of the evidence provided by the data in favor of
the hypothesis on the shape of the regression function, in addition to
its use in estimation.

This paper is organized as follows. Section \ref{sec2} presents the
Bernstein polynomial geometry and the hierarchical regression
model. Algorithms for\break Bayesian inference are given in the
\hyperref[app]{Appendix}. Section \ref{sec3} illustrates the method by simultaneously
analyzing all the data for these genes and indicates that this
method does bring insights into baculovirus biology. Section \ref{sec4}
concludes with a brief discussion.

\section{Bayesian inference}\label{sec2}

\subsection{Bernstein polynomial geometry}\label{sec2.1}

Let $F_{c,a}(t)=\sum_{i=0}^n
a_i\varphi_{i,n}(\frac{t-c}{1-c})I_{(c,1]}(t)$, where
$a=(a_0,\ldots,a_n)$. Proposition \ref{prop1} provides a sufficient
condition on $a$ under which $F_{c,a}$ is in $\mathcal A$.
Proposition \ref{prop2} complements Proposition \ref{prop1} and provides
Bernstein--Weierstrass type approximations for functions in
$\mathcal A$. In this paper derivatives at 0 and 1 are meant to
be one-sided. All the proofs of the propositions in this paper are
omitted, because they are similar to those in Chang et al. (\citeyear{Chaetal2005})
and Chang et al. (\citeyear{Chaetal2007}).

\begin{prop}\label{prop1}
Let $n \ge 3$ and $c\in [0,1).$ If $0=a_0=a_1\leq
\min_{l=2,\ldots,n}a_{l} < \max_{l=2,\ldots,n}a_{l}$,
then $F_{c,a}$ is continuously differentiable, constantly $0$ on
$[0,c],$ and larger than $0$ on $(c,1)$.
\end{prop}

Let $I_n=\{F_{c,a} \mid  c \in [0,1),a=(a_0,\ldots,a_n) $
satisfying $ 0=a_0=a_1\leq \break\min_{l=2,\ldots,n}a_{l} <
\max_{l=2,\ldots,n}a_{l}\}$.
For two continuously differentiable functions $f$ and $\tilde{f}$, define
$e(f,\tilde{f})=\|f-\tilde{f}\|_{\infty}+\|f'-{\tilde{f}}^{'}\|_{\infty}$,
where $f'$ denotes the derivative of $f$, and $\|\cdot\|_{\infty}$
is the sup-norm for functions on $[0,1]$. Then we have
the following:

\begin{prop}\label{prop2}
Let $\mathcal{D}=\bigcup^{\infty}_{n=3} I_n$. Then $\mathcal{D}$
is dense in $\mathcal A$, under $e$.
\end{prop}




\subsection{Bayesian regression model}\label{sec2.2}

\mbox{}

(i) \textit{Hierarchical prior}

For each $g=1,\ldots,154$, we will introduce probabilities $\pi_g$
on $\mathcal A$ as follows. We first describe the framework and
then the specific priors to be used. Let $\pi_{1,g}$ be a
probability density function on $[0,1]$, meant to be the prior on
the onset time $c$ of gene $g$; $\pi_{2,g}$ be a probability mass
function on the set of positive integers $\{3,4,\ldots \}$; for
each $n$, $\pi_{3,g}(\cdot\vert n)$ be a probability density
function on $\mathbb{R}^{n-1}$ of $b_n$. The probability
density/mass functions $\pi_{1,g}$, $\pi_{2,g}$ and $\pi_{3,g}$
jointly define a probability $\tilde{\pi}_g$ on $\mathcal{B}$ by
the product $\pi_{1,g}(c)\times \pi_{2,g}(n) \times
\pi_{3,g}(b_{n}\vert n)$; this in turn defines a probability
measure on $\mathcal A$ by (\ref{eq1.3}). Let $\pi_{4,g}$ be a probability
density on $\mathbb{R}^{1}$ for $\mu_g$, the mean of
$\epsilon_{jkg}$. Then $\pi_g=\tilde{\pi}_g\times \pi_{4,g}$ is
the prior density we will use on $\mathcal{B}\times\mathbb{R}^{1}$.

We now describe the strategies to specify $\pi_{1,g}$,
$\pi_{2,g}$, $\pi_{3,g}$ and $\pi_{4,g}$. Because our preliminary
studies based on a single gene suggest that the posterior
distributions of several features do not vary much with the prior
order of the Bernstein polynomial so long as it is not too small,
we take $\pi_{2,g}$ to have probability 1 for $n=15$, which has
the advantage of lessening the computational burden. The priors
$\pi_{1,g}$, $\pi_{3,g}$ and $\pi_{4,g}$ are defined in the
following by crude estimates based on all the 154 genes.

For each $g=1,\ldots,154$, let $\overline{Y}_{(0)g}\leq
\overline{Y}_{(1)g}\leq\cdots\leq\overline{Y}_{(15)g}$ be the
order statistics for \{$\overline{Y}_{0g}, \overline{Y}_{1g},
\ldots,\overline{Y}_{15g}$\}, where
$\overline{Y}_{kg}=\sum_{j=1}^{4} Y_{jkg}/4$. The prior
$\pi_{4,g}$ is the uniform distribution on
[0, $2\overline{Y}_{0g}$].

We now define $\pi_{1,g}$ for onset time. Let $\tilde k{(g)}$ be
the integer such that $\overline{Y}_{\tilde
k{(g)}g}=\overline{Y}_{(15)g}$; let $k{(g)}= \max \{ k   \mid
  k=0,1,\ldots,\tilde k{(g)} \mbox{ satisfying }
\overline{Y}_{kg} \leq 2\overline{Y}_{0g} \}+1$. Let $\tilde
X_{g}=X_{k{(g)}}$ and $\hat X_g$ equals $X_{(k{(g)}+\tilde
k{(g)})/2}$ if $(k{(g)}+\tilde k{(g)})/2$ is even, and equals
$X_{(k{(g)}+\tilde k{(g)}+1)/2}$ otherwise. Let $\alpha_1$ and
$\alpha_2$ be chosen so that the beta distribution
$\mathit{Beta}(\alpha_1,\alpha_2)$ has mean ${\sum_{g=1}^{154}
(\tilde X_{g}/\hat X_{g})/154}$ and variance
\[
\Bigl[ \Bigl(\max_{g=1,\ldots,154} \{\tilde
X_g/\hat X_{g} \}-\min_{g=1,\ldots,154} \{\tilde
X_g/\hat X_{g} \} \Bigr)\big/4 \Bigr]^2.
\]
Let $\phi_{11}={\alpha_1}-0.5$, $\phi_{12}={\alpha_1}+0.5$,
$\phi_{21}={\alpha_2}-0.5$ and $\phi_{22}={\alpha_2}+0.5$. We note
that for the present dataset, $\alpha_1=2.7771$ and
$\alpha_2=2.4481$, thus, $\phi_{11}=2.2771$, $\phi_{12}=3.2771$,
$\phi_{21}=1.9481$ and $\phi_{22}=2.9481$.

Let $\phi_1$ and $\phi_2$ be two random variables having
distributions respectively $\mathit{Uniform}(\phi_{11}, \phi_{12})$
and $\mathit{Uniform}(\phi_{21},\phi_{22})$. Let
$U_1,\ldots,U_{154}$ be a random sample of size 154 such that the
conditional distribution of $U_g$ given $\phi_1$ and $\phi_2$ is
$\mathit{Beta}(\phi_1,\phi_2)$ for each $g=1,\ldots,154$. We assume
that conditional on $\phi_1$ and $\phi_2$, the prior density
$\pi_{1,g}$ of the onset time of gene $g$ is the probability
density function of $\hat X_{g}\times U_g$. In particular, we
assume the onset time is in the interval $[ 0, \hat X_{g} ]$;
this assumption results from examining the data closely.

We next define $\pi_{3,g}(\cdot\vert n)$, which takes into
consideration the range of the observed expression levels and is
motivated by the propositions in Section \ref{sec2.1}. Let
$Y_{j[k']g}=Y_{jkg}$, if $\overline{Y}_{(k')g}=\overline{Y}_{kg}$.
Denote by $Y_{(1[k])g}\leq Y_{(2[k])g}\leq Y_{(3[k])g}\leq
Y_{(4[k])g}$ the order statistics of
\{$Y_{1[k]g},Y_{2[k]g},Y_{3[k]g},Y_{4[k]g}$\}. Let $\phi_3$ and
$\phi_4$ be two random variables having distributions respectively
$\mathit{Uniform}(\phi_{31}, \phi_{32})$ and
$\mathit{Uniform}(\phi_{41},\phi_{42})$, where $\phi_{31}$,
$\phi_{32}$, $\phi_{41}$ and $\phi_{42}$ are constants to be
assigned later. Let $V_{2,g},\ldots,V_{15,g}$ be a random sample
such that the conditional distribution of each $V_{i,g}$ given
$\phi_3$ and $\phi_4$ is $\it {Beta}$$(\phi_3,\phi_4)$. We assume
that conditional on $\phi_3$ and $\phi_4$, the prior density
function $\pi_{3,g}(\cdot \vert n)$ of the coefficients
$b_{n,g}=(b_{2,n,g},\ldots,b_{n,n,g})$ is the joint probability
density function of $2Y_{(4[15])g}\bullet
(V_{2,g},\ldots,V_{n,g})$. In the present study,
$\phi_{31}=\phi_{41}=0.5$ and $\phi_{32}=\phi_{42}=1.5$, which
give a large support of the prior. Let
$\phi=(\phi_1,\phi_2,\phi_3,\phi_4)$, which are the
hyperparameters.

Thus, under the assumption that $(c_g,n,b_{n,g},\mu_g)\in\mathcal{B}\times\mathbb{R}^{1}$ are conditionally independent given
$\phi$, the posterior density $\nu$ of all the parameters and
hyperparameters, given the data, is proportional to
\begin{equation}\label{eq2.1}
\qquad\Biggl\{\prod^{154}_{g=1}\prod^{K}_{k=0}
\prod^{m_k}_{j=1}\tilde
g_{kg}\bigl(Y_{jkg}-F_{c_g,b_{n,g}}(X_k)\bigr)\pi_g(c_g,n,b_{n,g},\mu_g\vert \phi)\Biggr\}\times
\psi( \phi),
\end{equation}
where $\tilde g_{kg}$ is the normal
density of $\epsilon_{jkg}$ specified in (\ref{eq1.2}) and
$\psi(\phi)=\prod^{4}_{i=1}(\phi_{i2}-\phi_{i1})^{-1}$ is the
joint hyperprior density function.

(ii) \textit{Sampling the posterior distributions}

Based on the hierarchical model, we use a Metropolis-within-Gibbs
algorithm to generate the posterior distributions for inference;
details of the algorithm are in the \hyperref[app]{Appendix}. The software is
written in Matlab, which is available from the author upon
request. The variance $\sigma_{kg}^{2}$ in (\ref{eq1.2}) to be used in the
algorithm is decided as follows. Let
$\tilde\sigma_{kg}^{2}=\sum_{j=1}^{4}(Y_{jkg}-\overline{Y}_{kg})^{2}/3$
and $\hat\xi_g\in \{0,1,2\}$ be the number that minimizes
$L(\xi_g)=\sum_{k=0}^{15}(Q_{kg}-\overline Q_g)^2/15$ with
$Q_{kg}=\tilde\sigma_{kg}^{2}/\overline{Y}_{kg}^{\xi_g}$ and
$\overline Q_g=\sum_{k=0}^{15}Q_{kg}/16$ for $\xi_g=0,1$ and 2.
With $x^{(t)}$ denoting the current state of the Markov chain and
$\hat \mu_g$ the background noise in the current state $x^{(t)}$,
we use
\[
\hat\sigma_{kg}^{2} =\hat\sigma_g^{2}\bigl(\hat F_g(X_{k})+\hat \mu_g\bigr)^{\hat\xi_g}
\]
for the $\sigma_{kg}^{2}$ in (\ref{eq1.2}) when updating $x^{(t+1)}$,
where $\hat\sigma_g^{2}=\sum_{k=0}^{15}(\tilde
\sigma_{kg}^{2}/\small{\overline{Y}}_{kg}^{\hat\xi_g})/16$ and
$\hat F_g$ is the $F_g$ determined by $x^{(t)}$.

We run 5 MCMC chains with initial values chosen randomly from the
hyperpriors and the priors of each gene $g$, and monitor
convergence by the Gelman--Rubin statistic $\hat R$, following the
suggestion in Gelman and Rubin (\citeyear{GelRub1992}) and Gelman et al. (\citeyear{Geletal2004}),
pages 294--297. For each of the 154 genes, the Gelman--Rubin
statistics $\hat R$ is calculated for six estimands of interest,
which are onset time (Ton), time to maximum (Tmax), maximum (Max),
time at which the slope is the highest (\mbox{Tslope}), the highest slope
(Slope) and the area under the curve on $[0,1]$. Each of the five
chains is run with 126,000,000 MCMC iterations and with a burn-in
period of 12,600,000 iterations, in which almost all the $\hat R$
are less then 1.1. The 56,700 updates, collected by taking one for
every 10,000 updates in the last 90\% of updates of these 5
sequences, are considered the sample from the posterior distribution,
which form the
basis for inference.

(iii) \textit{Numerical performance}

\begin{figure}[b]

\includegraphics{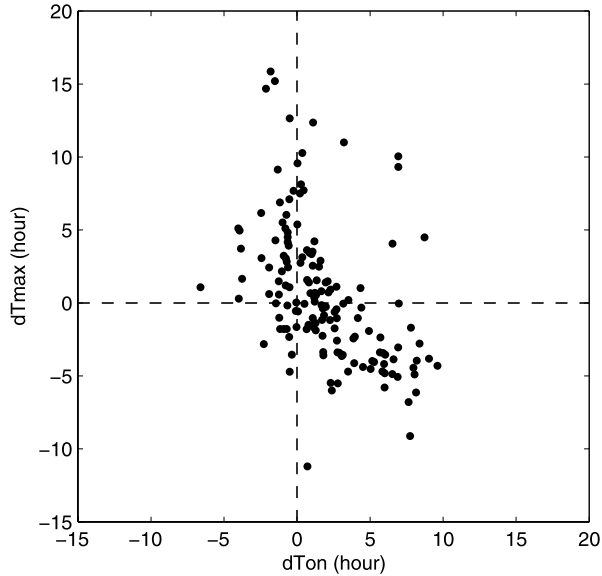}

  \caption{The differences between the Ton (Tmax)
based on the hierarchical Bayesian method and the crude estimate.
The first (second) coordinate of a dot is the onset time (time to
maximum) of a gene obtained from hierarchical Bayesian method
minus that of the same gene using naive method.}\label{fig1}
\end{figure}

To evaluate the numerical performance of the above hierarchical
Bayesian method, we studied a similar, but not hierarchical,
Bayesian method for the analysis of the time course expression of
a single virus gene.  This nonhierarchical Bayesian method,
modeling the expression profile also by Bernstein polynomials, is
more flexible in the sense that it allows nontrivial prior
probability on the order of the Bernstein polynomial and is
amenable to simulation studies.  In fact, the simulation studies
indicate its excellent numerical performance. Details of this
method and the simulation studies are in the supplementary article
[Chang et al. (\citeyear{Chaetal2008})]. We will evaluate the performance of the
hierarchical Bayesian method by comparing it with that of the nonhierarchical Bayesian method, in the context of analyzing our
baculovirus expression data.  The genes that we chose to conduct
this evaluation are selected by the criterion described in the
following paragraph; this choice serves also the purpose of
comparing the results from our hierarchical Bayesian method and
that in Jiang et al. (\citeyear{Jiaetal2006}), in addition to evaluating the
numerical performance of our method.

\begin{table}
\tablewidth=270pt
\caption{Estimates of the onset time based on the naive estimate,
hierarchical Bayesian method and Bayesian method}\label{tab1}
\begin{tabular*}{270pt}{@{\extracolsep{\fill}}lccccc@{}}
\hline
                  &\textbf{Jiang et al.}  &  \multicolumn{2}{c}{\textbf{Hierarchical}} \\
                  &      \textbf{(\citeyear{Jiaetal2006})}  &  \multicolumn{2}{c}{\textbf{Bayesian}}  &  \multicolumn{2}{c@{}}{\textbf{Bayesian}}    \\[-5pt]
&\multicolumn{1}{c}{\hrulefill}&\multicolumn{2}{c}{\hrulefill}&\multicolumn{2}{c@{}}{\hrulefill}\\
         \textbf{ID (Name)}&         \textbf{Estimate}   &                  \textbf{Mean}  & \textbf{Stdv}                 &       \textbf{Mean}  & \textbf{Stdv}                \\
         \hline
ID 130 ($\mathit{p10}$)&    0.0697          &                0.0724  &  0.0080  &   0.0756  &  0.0043             \\
ID 143 ($\mathit{pe38}$)&   0.0335          &                0.0293  &  0.0110  &   0.0211  &  0.0119               \\
ID 145 ($\mathit{pk}$-$\mathit{1}$)&   0.1785          &                0.1931  &  0.0025  &   0.1930  &  0.0025                   \\
ID 146 ($\mathit{pk}$-$\mathit{2}$)&   0.0552          &                0.0349  &  0.0091  &   0.0292  &  0.0104                     \\
ID 152 ($\mathit{v}$-$\mathit{cath}$)& 0.1374          &                0.1836  &  0.0072  &   0.1802  &  0.0095                  \\
\hline
\end{tabular*}
\end{table}

\begin{table}[b]
\tablewidth=260pt
\caption{Estimates of the time to maximum based on the naive
estimate, hierarchical Bayesian method and Bayesian method}\label{tab2}
\begin{tabular*}{260pt}{@{\extracolsep{\fill}}lccccc@{}}
\hline
                        &\textbf{Jiang et al.}  &  \multicolumn{2}{c}{\textbf{Hierarchical}} &  \\
                             &   \textbf{(\citeyear{Jiaetal2006})}    &  \multicolumn{2}{c}{\textbf{Bayesian}}     &  \multicolumn{2}{c@{}}{\textbf{Bayesian}}   \\[-5pt]
&\multicolumn{1}{c}{\hrulefill}&\multicolumn{2}{c}{\hrulefill}&\multicolumn{2}{c@{}}{\hrulefill}\\
                    \textbf{ID (Name)}&         \textbf{Estimate}   &                  \textbf{Mean}  & \textbf{Stdv}                 &       \textbf{Mean}  & \textbf{Stdv}                \\
\hline
           ID 130 ($\mathit{p10}$)&     0.7343         & 0.5855     & 0.0079   &  0.5859      &   0.0068            \\
           ID 143 ($\mathit{pe38}$)&    0.2185         & 0.3536     & 0.0248   &  0.3479      &   0.0230              \\
           ID 145 ($\mathit{pk}$-$\mathit{1}$)&    0.3515         & 0.5293     & 0.0046   &  0.5285      &   0.0051                    \\
           ID 146 ($\mathit{pk}$-$\mathit{2}$)&    0.2127         & 0.4163     & 0.0179   &  0.4171      &   0.0166                   \\
           ID 152 ($\mathit{v}$-$\mathit{cath}$)&  0.3564         & 0.4990     & 0.0082   &  0.4924      &   0.0101                \\
\hline
\end{tabular*}
\end{table}

For each gene, we consider the differences between the times
obtained from the hierarchical Bayesian method and those in Jiang
et al. (\citeyear{Jiaetal2006}).  Figure \ref{fig1} gives a rough idea of the differences.
The first (second) coordinate of a dot in Figure~\ref{fig1} is the onset
time (time to maximum) of a gene obtained from the hierarchical
Bayesian method minus that of the same gene using the naive method.  A
gene is selected if either its difference in onset times or that
in times to maximum is larger than 10 hours; we note that a
difference of this size may cause concerns in biological
interpretation. There are in total five such genes and their
differences in onset times are not as large as their differences
in the time to maximum; we carry out time course expression for
these five genes separately by the nonhierarchical Bayesian
method. The onset times and the times to maximum of these five
genes are shown in Table \ref{tab1} and Table \ref{tab2} respectively.  The first
column of Table \ref{tab1} gives the ID and the name of these genes; column
2 gives the onset times from Jiang et al. (\citeyear{Jiaetal2006}); column 3 gives
the means and standard deviations (Stdv) of the posterior
distributions of the onset times from the hierarchical Bayesian
method; column~4 gives those from the nonhierarchical Bayesian
method.  The entries in Table \ref{tab2} bear similar meanings as those in
Table \ref{tab1}.  It is clear from these tables that the results from the
hierarchical Bayesian method and those from the nonhierarchical
Bayesian method are in quite good agreement.   This suggests that
the hierarchical Bayesian method seems to produce reliable results in
the study of baculovirus gene expression.

We note that one of the genes, $ph$, was knocked out and we
included it in the hierarchical Bayesian analysis as a way to see
if our method is capable of  identifying it. Indeed, it does; it has
its time course expression profile much lower than all the others;
details are omitted. We also note that we compared other features
of several genes obtained from the hierarchical Bayesian method
and those from the nonhierarchical Bayesian method and find them
in very good agreement. To shorten the paper, we do not report the comparison.

One referee raised the question of whether our procedure
automatically identifies genes having different shapes like the
two singled out by initially examining the data. Indeed, based on
the posterior distributions, we get these two genes identified by
performing posterior predictive checking, as described in Gelman
(\citeyear{Gel2003}) and Gelman, Meng and Stern (\citeyear{GelMenSte1996}).

\section{Applications to the baculovirus data}\label{sec3}

Based on the samples from the posterior distribution obtained in
Section \ref{sec2}, this section carries out a genome-wide expression
analysis of the baculovirus and compares the results with those in
Jiang et al. (\citeyear{Jiaetal2006}).  It seems that the method of this paper
reveals more insights into virus biology than the naive method and in
case the results from this paper and those in Jiang et al. (\citeyear{Jiaetal2006})
are significantly different, it is more often than not that the
results from this paper are in better agreement with biology.
Since one of the genes, $ph$, was knocked out, the analysis in
Jiang et al. (\citeyear{Jiaetal2006}) was based on 155 genes and the following
studies regard the expression of the 153 genes.

\subsection{Times to maximum}\label{sec3.1}

According to Table \ref{tab2}, the differences in times to maximum for 5
genes are larger than ten hours. Except for the gene $\it p10$,
our method gives larger times to maximum. The following comments
seem to suggest that the times to maximum from the current approach
allow better or equally sensible interpretation, based on their
gene product function.

$\mathit{pe38}$ encodes a transcription factor important for virulence
of the baculovirus [Milks et al. (\citeyear{Miletal2003})]. It was shown that it expresses
from the immediate early phase throughout the late phase [Knebel-Morsdorf
et al. (\citeyear{KneMoretal1996})].  Larger time to maximum might reflect this fact more
satisfactorily.

$\mathit{pk}$-$\mathit{1}$ is a component of Ac\textit{M}NPV very late gene
transcription complex [Mishra, Chadha and Das (\citeyear{MisChaDas2008})].  Reilly and Guarino
(\citeyear{ReiGua1994}) indicated that the transcription of $\mathit{pk}$-$\mathit{1}$ peaks
in the very late stage of the infection cycle. Larger time to maximum
seems more consistent with these observations. Although there is
no report on the transcription time of $\mathit{pk}$-$\mathit{2}$, we tend
to think that it is similar to $\mathit{pk}$-$\mathit{1}$ and hence
transcribes also in the late stage of the infection cycle.

$v$-$\mathit{cath}$ encodes a papain type cysteine proteinase with
cathapsin L-like property.  Its proteinase activity is required
for the breakdown of host tissues during the later stages of virus
infection/pathogenesis [Hill, Kuzio and Faulkner (\citeyear{HilKuzFau1995})].  Larger time to maximum
better reflects the needs for its protein expression during this
stage, when the host has been exhausted completely and the virus can
be spread to other hosts most efficiently.

For the well-known late gene $\mathit{p10}$, although the hierarchical
Bayesian method gives a smaller time to maximum than that in Jiang
et al. (\citeyear{Jiaetal2006}), we note that this smaller time to maximum is still
the third largest among all the times to maximum of the 153 genes
and hence seems to cause less concern.

\subsection{Time course expression analysis}\label{sec3.2}

\begin{table}[b]
\caption{Data analysis for the gene $v$-$\mathit{cath}$}\label{tab3}
\legend{Table 3a. Posterior probability (Po), prior probability (Pr), the ratio of Po to Pr, and the Bayes factor (Bf) of being unimodal on $[0,\tau]$.}
\begin{tabular}{@{}lccc@{}}
\hline
                        $\bolds{[0,\tau]}$    &  \textbf{[0, 0.6667]} & \textbf{[0, 0.8333]} & \textbf{[0, 1.0000]}           \\
\hline
                        Po            &    1.0000    &    0.3280   &  0.0280         \\
                        Pr            &    0.4158    &    0.2658   &  0.0972              \\
                        Po$/$Pr         &    2.4050    &    1.2340   &  0.2881              \\
                        Bf            &    $\infty$  &    1.3482   &  0.2676          \\
\hline
\end{tabular}

\legend{Table 3b. Posterior probability (Po), prior
probability (Pr), the ratio of Po to Pr, and the Bayes factor (Bf)
that it is increasing before reaching its global maximum.}
\begin{tabular}{@{}lc@{}}
\hline
                        Po        &        1.0000               \\
                        Pr        &        0.3719                \\
                        Po$/$Pr     &        2.6889                 \\
                        Bf        &       $\infty$                \\
\hline
\end{tabular}

\end{table}
\begin{table}
\legend{Table 3c. The Ton, Tmax, Max, Tslope, Slope,
$L_{1}$-norm and Tend of the mode of the posterior density $\nu$
in (\ref{eq2.1}) is given in the third column in the table. The sample
mean, sample Stdv and support of the posterior probability
distribution and the prior probability distribution of these
features are respectively given in the fourth, fifth and sixth
column.}
\begin{tabular}{@{}llcccc@{}}
\hline
\textbf{Estimand} &                   &\textbf{Mode}      &   \textbf{Mean}     &  \textbf{Stdv}  &  \textbf{Support}            \\
\hline
                            Ton   &         Posterior & 0.1819   & 0.1836     & 0.0072 &  (0.1197, 0.2079)   \\
                                  &         Prior     &          & 0.1329     & 0.0510 &  (0.0023, 0.2498)  \\[5pt]
                            Tmax  &         Posterior & 0.5093   & 0.4990     & 0.0082 &  (0.4444, 0.5231)    \\
                                  &         Prior     &          & 0.7902     & 0.2278 &  (0.2083, 1.0000)    \\[5pt]
                            Max   &         Posterior & 1.7779   & 1.6797     & 0.0877 &  (1.2713, 1.9397)  \\
                                  &         Prior     &          & 2.1139     & 0.5189 &  (0.3668, 3.0793)    \\[5pt]
                          Tslope  &         Posterior & 0.2176   & 0.2358     & 0.0420 &  (0.1944, 0.4074)    \\
                                  &         Prior     &          & 0.4236     & 0.3499 &  (0.0509, 1.0000)    \\[5pt]
                          Slope   &         Posterior & 8.1613   & 8.7394     & 1.0778 &  (5.6079, 13.5625)  \\
                                  &         Prior     &          & 15.0351    & 9.1052 &  (1.7144, 59.4057)  \\[5pt]
                     $L_{1}$-norm &         Posterior & 0.6206   & 0.6005     & 0.0298 &  (0.5004, 0.7357)   \\
                                  &         Prior     &          & 1.0878     & 0.3419 &  (0.1173, 2.2368)   \\[5pt]
                            Tend  &         Posterior & 0.8380   & 0.8400     & 0.0720 &  (0.7500, 1.0000)   \\
                                  &         Prior     &          & 0.9366     & 0.1217 &  (0.3611, 1.0000)   \\
\hline
\end{tabular}
\end{table}

\begin{figure}

\includegraphics{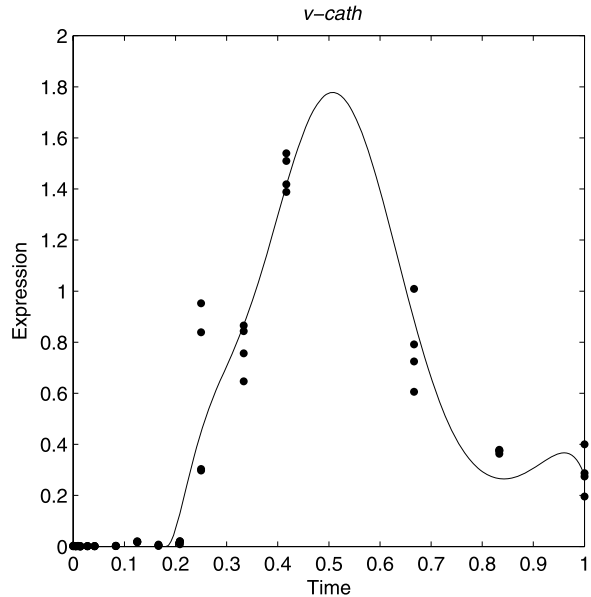}

  \caption{The data and the posterior mode of the time
course expression of the gene $v$-$\mathit{cath}$.}\label{fig2}
\end{figure}

To illustrate the use of our method, we now present, in Table \ref{tab3},
the features of the expression profile of the gene $v$-$\mathit{cath}$,
which is one of the genes selected to evaluate the numerical
performance of our method. Figure \ref{fig2} presents the data and the
posterior mode of its time course expression. Most of these
features can not be reliably obtained by the naive method. This
illustration also helps to appreciate that the data have
substantial contribution in the inference on these features of
$v$-$\mathit{cath}$.  Table \ref{tab3}a reports the posterior probability and the
prior probability that the parameter represents a unimodal curve
on the interval $[0,\tau]$ for $\tau = 0.6667,0.8333,1.0000$; the
last two rows give respectively the ratio of the posterior
probability to the prior probability and the Bayes factor.  Table
\ref{tab3}a presents  strong evidence, provided by the data, in favor of
the unimodality of the time course profile. The posterior
probability and the prior probability that the parameter
represents a curve that is increasing before reaching its global
maximum are reported in Table \ref{tab3}b; similarly, the last two rows
give respectively the ratio of the posterior probability to the
prior probability and the Bayes factor; Table \ref{tab3}b strongly suggests
that the expression profile increases before its global maximum.

Let $\tau_0$ (Tend) denote the largest time point $t$ such that
the time course expression profile is unimodal on $[0,t]$. Let
$L_1$-norm denote the area under the time course expression
profile on $[0,\tau_0]$. Table \ref{tab3}c reports Ton, Tmax, Max, Tslope,
Slope, $L_1$-norm and Tend of the mode of the posterior density
$\nu$ in (\ref{eq2.1}) and the sample mean, sample standard deviation
(Stdv) and support of these features on the sample respectively
from the posterior and prior distributions. Comparing the Stdv and
the support from the posterior and the prior, we know that the
data have substantial contribution in the inference on these
features.

\begin{figure}

\includegraphics{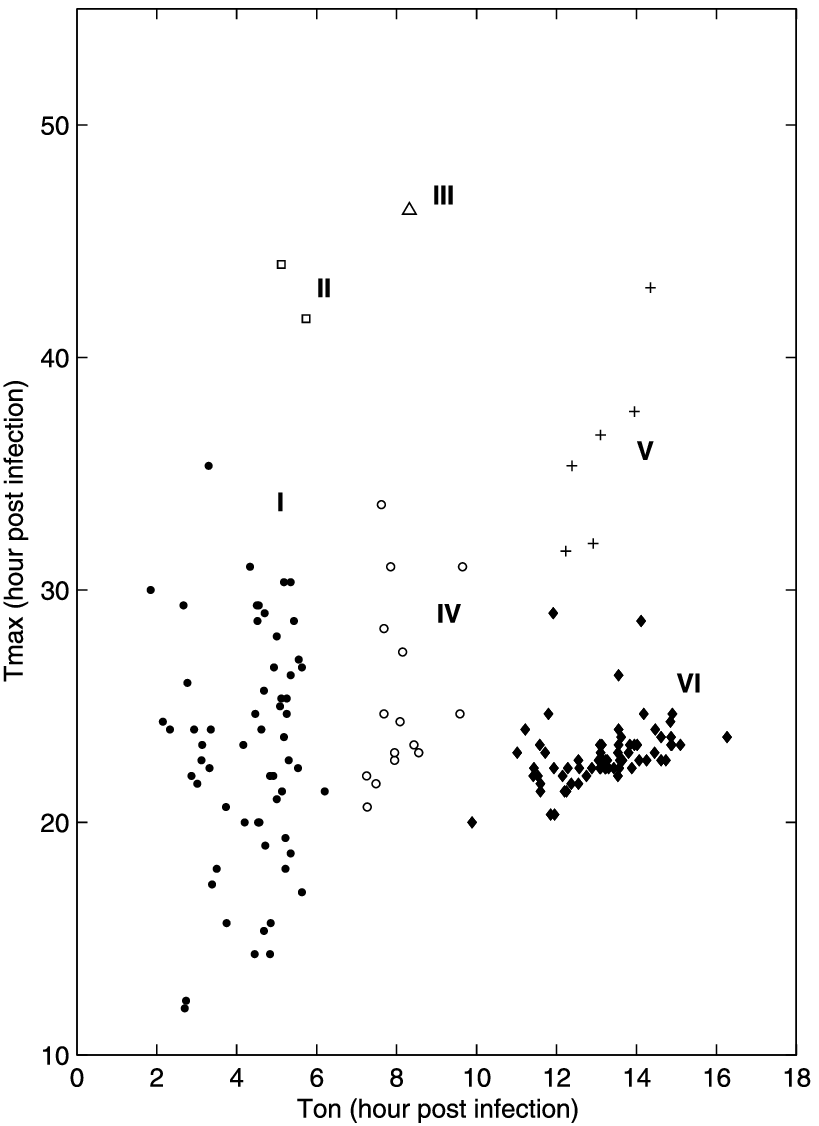}

\begin{longlist}
\item[Group I (early onset and early to maximum)]
\textit{35K/p35, egt, me53, 39K/pp31, pcna, 94K, ie-2, lef1,
    pnk/pnl, he65, ie-01, ie-1, lef6,  pk-2, DNA-pol, gp64,
    pe38, lef3, p48, lef7, p26, ctx, helicase,
    lef11, lef2, p15, tlp, orf-603}
\item[Groups II and III (mid-course onset and late to maximum)]
\textit{orf-1629, p10, p74}
\item[Group IV (mid-course onset and early to maximum)]
\textit{gta, p40, ptp, iap1, p43, alk-exo, cg30, odv-e18, PE/pp34}
\item[Group V (late onset and mid-course to maximum)]
\textit{pk-1, v-cath}
\item[Group VI (late onset and early to maximum)]
\textit{gp41, p47, p6.9, vlf-1, chitinase, ie-0, pkip, sod,
    lef9, odv-ec27, lef5, env-prot, lef4, lef8, p95, vp39,
    gp16, 38K, bro, fgf, fp, HisP, iap2, odv-e56,
    v-ubi, 49K, odv-e25, vp80, gp37, lef10, p24, odv-e66}
\end{longlist}
  \caption{A classification methodology for the 153 genes based on
Ton and Tmax. Selected known genes in each classified group are
listed at the bottom.}\label{fig3}
\end{figure}

It is customary in microarray literature to cluster genes
according to their expression profiles for biologists to use.
Using the Ton and Tmax of the mode of the posterior distribution,
we apply the cluster analysis algorithm proposed by Hall and
Heckman (\citeyear{HalHec2002}) to cluster the 153 genes into six groups, which are
I (early onset and early to maximum), IV (mid-course onset and
early to maximum), V~(late onset and mid-course to maximum), VI
(late onset and early to maximum), and II and III (mid-course
onset and late to maximum). The scatterplot in Figure \ref{fig3} reports
the cluster analysis result; genes with known functions are listed
according to the clusters to which they belong.

While Figure \ref{fig3} helps to shed light on the gene groups, it would be
interesting to see if genes in the same group have a more similar
overall expression profile. Using the rank correlation of two time
course expression profiles as the distance between two genes,
Table \ref{tab4} shows that the means of the rank correlation for two genes
randomly chosen from the same one of the clusters are smaller than
that from the set of all 153 genes. We note that the rank
correlation is a measure of similarity between functions studied
by Heckman and Zamar (\citeyear{HecZam2000}). This seems to suggest that genes in
the same group have a more similar expression profile.

\begin{table}
\tablewidth=230pt
\caption{Mean and standard deviation of the rank correlation of
the time course expression of two genes chosen from specific
groups}\label{tab4}
\begin{tabular*}{230pt}{@{\extracolsep{\fill}}lccc@{}}
\hline
&  &  \multicolumn{2}{c@{}}{\textbf{Rank correlation}}  \\[-6pt]
&&\multicolumn{2}{c@{}}{\hrulefill}\\
                    \textbf{Group}&      \textbf{Number of genes}   &              \textbf{Mean}  & \textbf{Stdv}                       \\
\hline
                         \phantom{II}I &     \phantom{0}60                 & 0.8070             & 0.1578           \\
                        \phantom{I}II &      \phantom{00}2                 & 0.9793             & 0.0000                \\
                       III &      \phantom{00}1                 & NA\tabnoteref{table1}           & NA          \\
                        IV &     \phantom{0}15                 & 0.8852             & 0.0807                      \\
                         \phantom{I}V &      \phantom{00}6                 & 0.9108             & 0.0594                    \\
                        VI &      \phantom{0}69                & 0.8981             & 0.0867                  \\
                      All  &      153               & 0.7717             & 0.2023                  \\
\hline
\end{tabular*}
\tabnotetext{table1}{NA means not applicable.}
\end{table}

\begin{figure}

\includegraphics{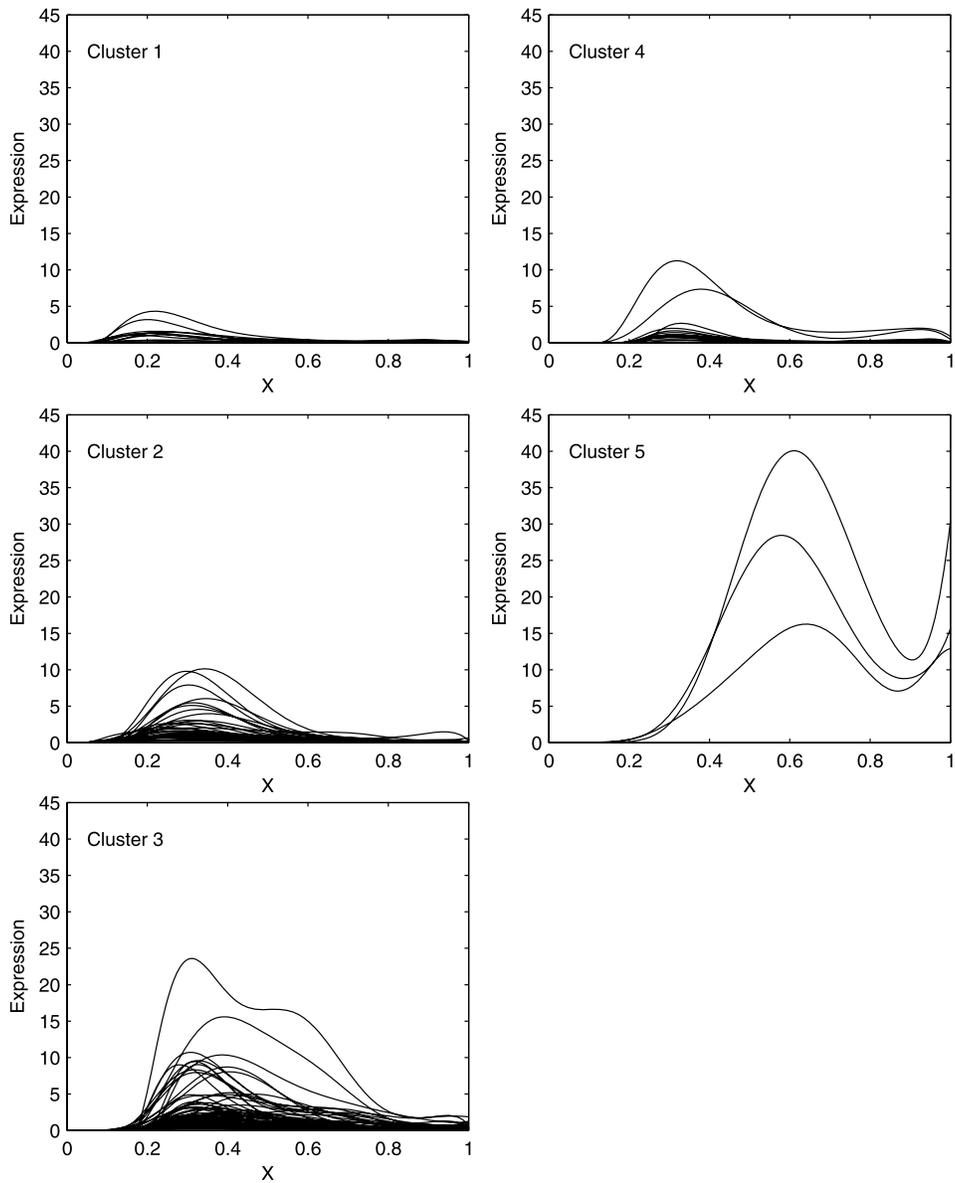}

  \caption{Cluster analysis for the 153 viral gene
expression profiles.}\label{fig4}
\end{figure}

Based on the time course expression profile of the 153 genes
obtained by the posterior mode, we use the $K$-means algorithm along
with the sample rank correlation matrix to cluster them; as in
Jiang et al. (\citeyear{Jiaetal2006}), we also consider five clusters. The five gene
clusters are contained in Figure \ref{fig4}.

We note that clustering is an important step toward gaining
insights from high-throughput expression data and there is usually
some arbitrariness in forming clusters.  Since clustering in
Figure \ref{fig3} is based only on onset times and times to maximum, it is
easier to cluster and to interpret, but Figure \ref{fig4} is more
informative in general.  For example, Cluster 5 in Figure \ref{fig4}
consists of three genes; one of the most obvious features of these
three genes seems to be their large expression levels; thus, it is
interesting to note that they are also in such close proximity to
each other in Figure \ref{fig3} and they form exactly the Groups II and III
in Figure \ref{fig3}.

\subsection{Total expression amount and structure genes}\label{sec3.3}

It is of great interest to study the widely discussed conjecture
that the virus has a great demand of structural proteins. While we cannot provide a definitive answer to this question, we think the
method of this paper can shed some light on it. One of the salient
features of the expression profile obtained by our method is the
area under the time course expression profile ($L_1$-norm);
roughly speaking, the $L_1$-norm of a gene is the sum of the lives
of all the mRNA molecules transcribed during the time interval
ended at Tend; the life of an mRNA molecule is the time length
from its transcription to its degradation or its Tend. Although
the relation between  the \mbox{$L_1$-norm} and the total number of the
proteins translated is complex, we expect they are positively
correlated. We indicate in the following that structure genes seem
to have larger $L_1$-norms.  There are 74 baculovirus genes with
 known names, in which 15 of them are structure genes and the rest
are not.  We find that, in terms of  the \mbox{$L_1$-norm}, four of the five
largest genes are structure genes, giving an odds ratio of 21.1;
among the ten largest genes, five of them are structure genes,
giving an odds ratio of 5.4; among the 20 largest genes, 7 of them
are structure genes, giving an odds ratio of~3.1.  We also study by
the Wilcoxon statistic the null hypothesis that there is no difference
in the $L_1$-norm between structural genes and nonstructural
genes. We find the statistic has value 1.73 and using the one-sided Wilcoxon test, it has $p$-value 0.0418. This seems to
reinforce the conjecture that structural genes tend to have larger
$L_1$-norms. We note it seems hard to estimate the $L_1$-norms and
to study this conjecture by the method of Jiang et al. (\citeyear{Jiaetal2006}).

\subsection{Motif and onset time}\label{sec3.4}

Biologists tend to think that genes participating in the same
biological process may be transcriptionally coregulated.  One
preliminary step in studying this phenomenon might be to examine
whether upstream sequence motifs of a gene have something to do
with its transcription time. In the baculovirus literature [Ayres et
al. (\citeyear{Ayeetel1994}) and Friesen and Miller (\citeyear{FriMil2001}), for example], motifs
A(A/T)CGT(G/T) and CGTGC are called the early motif; motif TAAG is
called the late motif; genes having motif CATG  are usually believed
to express early.  Jiang et al. (\citeyear{Jiaetal2006}) studies this by reporting
the proportions of these motifs in the 5 gene clusters obtained
from clustering the time course expression crude data. While we
can conduct a similar study by means of the clusters obtained from
our Bayesian method, we propose to ignore the clusters and take a
more direct and relevant approach to address this issue.

\begin{table}
\tablewidth=320pt
\caption{Motifs have to do with onset time. Comparing the onset
times of genes having specific motifs with those without by the
Wilcoxon statistic, which is asymptotically standard normal. Minus
(plus) values indicate the former (latter) is smaller (larger)}\label{tab5}
\begin{tabular*}{320pt}{@{\extracolsep{4in minus 4in}}lcccc@{}}
\hline
\textbf{Motif}        &  \textbf{With}    & \textbf{Without} & \textbf{Wilcoxon statistic}& \textbf{$\bolds{p}$-value} \\
\hline
Early\tabnoteref{table2}  &    \phantom{0}64    &   66    &  $-2.65$ &   0.00402      \\
TAAG         &    \phantom{0}70    &   60    &  \phantom{$-$}4.04  &  0.00003        \\
CATG         &    \phantom{0}69    &   61    &  $-2.66$ &   0.00391           \\
Early/CATG   &    110   &   20    &  $-2.54$ &   0.00554       \\
\hline
\end{tabular*}
\tabnotetext{table2}{The early motif (Early) consists of motifs A(A/T)CGT(G/T) and  CGTGC.}
\end{table}

Based on the onset times of this paper, we study the hypotheses
that, with a given motif, there is no difference between the onset
times of the genes with this motif and those without this motif.
We study them by the Wilcoxon statistic.  Table~\ref{tab5} summarizes the
numbers of genes having or not having these motifs and reports the
Wilcoxon statistics and their $p$-values for testing the
corresponding one-sided null hypothesis.  For example, the second
row shows that 70 genes have TAAG and 60 genes do not have it, its
Wilcoxon statistic is 4.04 and the $p$-value is smaller than 0.0001,
which seem to suggest that the genes having TAAG tend to have
later onset times. It seems Table \ref{tab5} supports the idea that motifs
have something to do with onset times.

\subsection{Colocalization}\label{sec3.5}

\begin{figure}

\includegraphics{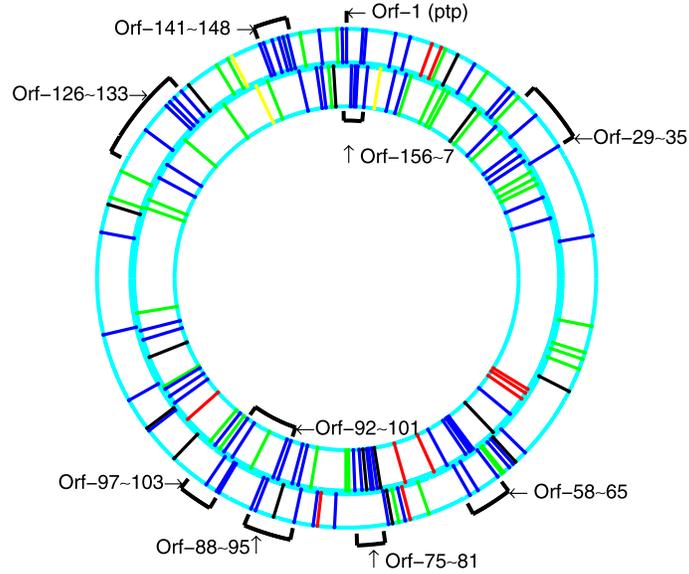}

  \caption{Genome map view of the five gene
clusters color tagged in the baculovirus genome. Red, green, blue,
black and yellow represent respectively genes in the cluster 1, 2,
3, 4, 5 in Figure \protect\ref{fig4}.}\label{fig5}
\end{figure}

Because functionally correlated or coregulated genes in an operon
of a bacterial genome may be located in nearby loci of the
physical genome [Lagreid et al. (\citeyear{Lagetal2003})], Jiang et al. (\citeyear{Jiaetal2006})
investigated whether a similar gene organization exists in the
Ac\textit{M}NPV genome. Based on the time course expression normalized
data, Jiang et al. (\citeyear{Jiaetal2006}) clustered genes into five clusters and
noted six colocalized clusters. A colocalized cluster is defined
as a genome region that contains at least five consecutive genes
from the same gene cluster where no more than one interruption
occurs by a gene from other gene clusters in either the plus or
minus strand. Using the same definition of a colocalized cluster,
we find there are nine colocalized clusters, based on the five
clusters exhibited in Figure \ref{fig4}. These nine colocalized clusters
are shown in Figure \ref{fig5}. This seems to suggest that expression
profiles from our sophisticated method reveals more signals than
the naive method.

The phenomenon that genes with similar expression profile tend to
be located near each other is referred to as colocalization in
Jiang et al. (\citeyear{Jiaetal2006}). Since the above definition of a colocalized
cluster is somewhat arbitrary, we present a more systematic study
on this in Table \ref{tab6}. Column two and column three of Table \ref{tab6} give
respectively the probability of two (three, four, five) randomly
chosen genes that belong simultaneously to the same one of the
five clusters and the probability of two (three, four, five)
randomly chosen neighboring genes that belong simultaneously to
the same one of the five clusters. Because the numbers in column 2
are smaller than those in column 3, it seems that colocalization
does exist.

\begin{table}[b]
\tablewidth=200pt
\caption{The probability that $N$ randomly chosen (neighboring)
genes belong to the same cluster in Figure \protect\ref{fig4}}\label{tab6}
\begin{tabular*}{200pt}{@{\extracolsep{\fill}}lcc@{}}
\hline
$\bolds{N}$         &  \textbf{Randomly chosen} & \textbf{Neighboring}           \\
\hline
2            &  0.3835      & 0.4837               \\
3            &  0.1820      & 0.2680                   \\
4            &  0.0926      & 0.1373                   \\
5            &  0.0484      & 0.0719               \\
\hline
\end{tabular*}
\end{table}

From the viewpoint of evolution, it might also be appealing to see
if genes close to each other on the genome have a similar expression
pattern. One relevant null hypothesis would be that there is no
difference in the rank correlation of expression profiles from
nearby genes and that from far away genes. For integer $0\leq
Z\leq 76=(153-1)/2$, let $\operatorname{Nei}(g,Z)$ denote the set of genes
whose distance from gene $g$ is no larger than $Z$; here the
distance between two genes is the number of genes lying strictly
between them. Let $\operatorname{Rn}(Z)$ denote the set of rank correlations of
the time course expression profile of a gene $g$ and that of a
gene in $\operatorname{Nei}(g,Z)$. Let $\operatorname{RCn}(Z)$ denote the set of rank
correlations of the time course expression profile of a gene $g$
and that of a gene not in $\operatorname{Nei}(g,Z)$. In terms of this notation,
the null hypothesis becomes that there is no statistical
difference between $\operatorname{Rn}(Z_1)$ and $\operatorname{RCn}(Z_2)$. We studied the
hypothesis by the Wilcoxon statistic for many choices of $Z_1$ and
$Z_2$. Table \ref{tab7} reports the Wilcoxon statistics and their $p$-values
for testing the corresponding one-sided null hypothesis for
several choices of $Z_1$ and $Z_2$. It suggests that nearby genes
do have a higher chance to have a similar expression pattern.

\section{Discussion}\label{sec4}

We have illustrated a hierarchical Bayesian shape restricted
regression method for the inference on the genome-wide time course
expression of virus genes and, based on the profiles provided by
this method, we have examined salient features on the time course
expression curves, studied some hypotheses on and thus brought
insights into baculovirus biology. It is to be noted that our
method helps to formulate biological questions quantitatively so
as to make modern statistics methods applicable. Although we
looked at colocalization, the relation between upstream motifs and
onset times, and that between area under curve and gene function,
these are, nevertheless, preliminary investigations. Further studies
are needed to give a more complete account of these aspects of
the baculovirus.

In view of the facts that genome-wide expression studies of virus
genes are gaining popularity, all the previous works in this area
use at most crude statistics for biological interpretation, and
the existing discrepancies between the studies need to be
resolved, we think our method is useful not only in one single
expression study of virus genes but also in comparing these
studies, which would enhance our understanding of the gene
regulation network. We note that our method can be used to provide
comprehensive comparison of the time course transcription profiles
from different experiments when even their time points are not
identical, as long as there are enough of them to capture their
respective main features.

\begin{table}
\caption{Comparing the rank correlation of the time course
expression profiles from nearby genes and that from far away
genes. $\operatorname{Rn}(Z)$ is the set of rank correlations for genes having no
more than $Z$ genes lying between them; $\operatorname{RCn}(Z)$ is that for genes
having at least $Z$ genes lying between them}\label{tab7}
\begin{tabular*}{\textwidth}{@{\extracolsep{\fill}}llcc@{}}
\hline
$\bolds{\operatorname{Rn}(Z)}$  &  $\bolds{\operatorname{RCn}(Z)}$          & \textbf{Wilcoxon statistic}& $\bolds{p}$\textbf{-value}           \\
\hline
$\operatorname{Rn}(2)$  &  $\operatorname{RCn}(12)$                 &  8.81 &   0.0000      \\
$\operatorname{Rn}(4)$  &  $\operatorname{RCn}(14)$               &  5.82 &   0.0000           \\
$\operatorname{Rn}(6)$  &  $\operatorname{RCn}(16)$             &  5.09 &   0.0000           \\
$\operatorname{Rn}(8)$  &  $\operatorname{RCn}(18)$               &  2.72 &   0.0033       \\
$\operatorname{Rn}(10)$ &  $\operatorname{RCn}(20)$               &  2.01 &   0.0221      \\
$\operatorname{Rn}(12)$ &  $\operatorname{RCn}(22)$               &  0.84 &   0.2002           \\
$\operatorname{Rn}(14)$ &  $\operatorname{RCn}(24)$               &  1.25 &   0.1051           \\
$\operatorname{Rn}(16)$ &  $\operatorname{RCn}(26)$               &  2.31 &   0.0105       \\
$\operatorname{Rn}(18)$  & $\operatorname{RCn}(28)$               &  1.59 &   0.0558      \\
$\operatorname{Rn}(20)$  & $\operatorname{RCn}(30)$               &  0.23 &   0.4078           \\
$\operatorname{Rn}(22)$  & $\operatorname{RCn}(32)$               &  0.55 &   0.2913           \\
$\operatorname{Rn}(24)$  & $\operatorname{RCn}(34)$               &  1.45 &   0.0737       \\
$\operatorname{Rn}(26)$  & $\operatorname{RCn}(36)$               &  0.25 &   0.4014       \\
\hline
\end{tabular*}
\end{table}

As for future methodological development, we think the
Bernstein--Dirichlet prior of Petrone (\citeyear{Pet1999}) and the related
samplers are also useful in this context; studies in this line and
comparison with the approach in this paper deserve our attention.




\begin{appendix}
\section*{Appendix: Metropolis-within-Gibbs algorithm for the posterior}\label{app}

Let $B_{n}=\{b_{n} \in \mathbb{R}^{n-1} \dvtx F_{c,b_{n}} \in I_{n}$
for some $c\in[0,1) \}$. Denote
$(b_{2,n,g},\ldots,\break b_{n,n,g})=b_{n,g}$ by
$(a_{2,g},\ldots,a_{n,g})=a_g$. Let $
\mathbf{c}=(c_1,\ldots,c_{154})$;
$\mathbf{a}=(a_1,\ldots,a_{154})$;
$\mathbf{u}=(\mu_1,\ldots,\mu_{154})$.

Let $B=\{\phi, \mathbf{c}, \mathbf{a}, \mathbf{u}
\mid \phi=(\phi_1,\phi_2,\phi_3,\phi_4)\in
[2.2771,3.2771]\times[1.9481,\break 2.9481]\times[0.5,1.5]\times[0.5,1.5],
c_g \in [0,\hat X_g], a_{g} \in B_n, \mu_g \in [0,2\overline
Y_{0g}] \}$. Our computational strategy consists of the following
five MCMC algorithms to update $\phi$, $\mathbf{c}$, $\mathbf{a}$
and $\mathbf{u}$ consecutively. Let
$x^{(t)}=(\phi^{(t)},\mathbf{c}^{(t)}, \mathbf{a}^{(t)},
\mathbf{u}^{(t)}) \in B$ denote the current state of the MCMC
chain for sampling the
posterior distribution.

(i) \textit{Update $\phi_1$ and $\phi_2$}

\begin{enumerate}
\item Let $\tilde \phi_1$ and $\tilde \phi_2$ be two random samples from
      $\mathit{Uniform}(\phi_{11},\phi_{12})$
      and\break $\mathit{Uniform}(\phi_{21},\phi_{22})$ respectively;
\item let $y=(\tilde \phi_1,\tilde \phi_2, \phi_3^{(t)},\phi_4^{(t)},\mathbf{c}^{(t)},
                \mathbf{a}^{(t)}, \mathbf{u}^{(t)})$;
\item set
\[
x^{(t+1)} = \cases{
         y  , & \quad with  prob. $\rho = \min  \biggl\{ 1, \displaystyle{\frac{\nu (y) }{{\nu (x^{(t)})  }}}  \biggr\}
         $,\cr
         x^{(t)}  , & \quad otherwise.
}
\]
\end{enumerate}

(ii) \textit{Update $\phi_3$ and $\phi_4$}

\begin{enumerate}
\item Let $\tilde \phi_3$ and  $\tilde \phi_4$ be two random samples from
        $\mathit{Uniform}(\phi_{31},\phi_{32})$ and\break $\mathit{Uniform}(\phi_{41},\phi_{42})$ respectively;
\item let $y=(\phi_1^{(t)},\phi_2^{(t)},\tilde \phi_3,\tilde \phi_4,\mathbf{c}^{(t)},
           \mathbf{a}^{(t)}, \mathbf{u}^{(t)})$;
\item set
\[
x^{(t+1)} = \cases{
         y,   &\quad with  prob. $ \rho = \min  \biggl\{ 1, \displaystyle{\frac{\nu (y) }{\nu (x^{(t)})  }}  \biggr\}
         $,\cr
         x^{(t)},   &\quad otherwise. }
\]
\end{enumerate}

(iii) \textit{Update} $\mathbf{c}$

There are 154 components ($c_1,\ldots,c_{154}$) in $\mathbf{c}$;
we update them one at a time in the order of the coordinates.
Suppose $c_1^{(t)},\ldots,c_{g-1}^{(t)}$ have been just updated
and we now want to update $c_g^{(t)}$.

\begin{enumerate}
  \item Let $U$ be a random sample from $\mathit{Beta}(\phi_1^{(t)},\phi_2^{(t)})$;
  \item let $\tilde c_g=\hat X_g\times U$; let $\pi_{1,g}(\tilde c_g
             \vert \phi_1^{(t)},\phi_2^{(t)})$ denote the prior density
             $\pi_{1,g}$ of $\tilde c_g$ given $\phi_1^{(t)}$ and
             $\phi_2^{(t)}$;
  \item let $y=(\phi^{(t)},{c}_1^{(t)},\ldots,{c}_{g-1}^{(t)},\tilde c_g,
             c_{g+1}^{(t)},\ldots,c_{154}^{(t)},
             \mathbf{a}^{(t)}, \mathbf{u}^{(t)})$;
  \item set
\[
x^{(t+1)} = \cases{
         y ,  & \quad with  prob. $\rho =
         \min  \biggl\{ 1, \displaystyle{\frac{\nu (y)\pi_{1,g}(c_g^{(t)}\vert \phi_1^{(t)},\phi_2^{(t)})  }{\nu (x^{(t)}) \pi_{1,g}(\tilde c_g \vert \phi_1^{(t)},\phi_2^{(t)})  }}  \biggr\}
         $,\cr
         x^{(t)},   & \quad otherwise.
}
\]
\end{enumerate}

(iv) \textit{Update} $\mathbf a$

We update one coordinate of $\mathbf a$ each time in the order of
the coordinates. Suppose we have updated
$a_{2,g}^{(t)},\ldots,a_{i-1,g}^{(t)}$ and we now want to update
$a_{i,g}^{(t)}$.

\begin{enumerate}
\item Let $V$ be a random sample from
             $\mathit{Beta}(\phi_3^{(t)},\phi_4^{(t)})$;
\item let $\tilde a_{i,g}=2Y_{(4[15])g}\times V$; let $\pi_{3,g}(\tilde
             a_{i,g}\vert \phi_3^{(t)},\phi_4^{(t)})$ denote the prior density
             $\pi_{3,g}(\cdot\vert n)$ of the coefficient $\tilde a_{i,g}$ given $\phi_3^{(t)}$ and
             $\phi_4^{(t)}$;

\item let $y$ be the same vector as $x^{(t)}$ except
  replacing $a_{i,g}^{(t)}$ by $\tilde a_{i,g}$;

\item set
\[
x^{(t+1)} = \cases{
         y,   & \quad with  prob. $\rho =
         \min  \biggl\{ 1, \displaystyle{\frac{\nu (y)\pi_{3,g}(a_{i,g}^{(t)}\vert \phi_3^{(t)},\phi_4^{(t)})
         }{\nu (x^{(t)}) \pi_{3,g}(\tilde a_{i,g}\vert \phi_3^{(t)},\phi_4^{(t)}) }}  \biggr\}
         $,\cr
         x^{(t)} ,  & \quad otherwise.}
\]
\end{enumerate}

(v) \textit{Update} $\mathbf u$

There are 154 components ($\mu_{1},\ldots,\mu_{154}$) in
$\mathbf{u}$; we update them one at a time in the order. Suppose
we have updated $\mu_1^{(t)},\ldots,\mu_{g-1}^{(t)}$ and we now
want to update $\mu_g^{(t)}$.

\begin{enumerate}
  \item Let $\tilde \mu_g$ be a random sample from $\mathit{Uniform}(0,2\overline Y_{0g})$;
  \item let $y=(\phi^{(t)}, \mathbf{c}^{(t)},
                  \mathbf{a}^{(t)},{\mu}_1^{(t)},\ldots,\mu_{g-1}^{(t)},\tilde \mu_g,
             \mu_{g+1}^{(t)},\ldots,\mu_{154}^{(t)})$;
  \item set
  \[
  x^{(t+1)} = \cases{
         y,   &\quad with  prob. $\rho =
            \min  \biggl\{ 1, \displaystyle{\frac{\nu (y) }{\nu (x^{(t)})  }}  \biggr\}
            $,\cr
         x^{(t)},   & \quad otherwise.
         }
  \]
\end{enumerate}
\end{appendix}

\section*{Acknowledgments}

We are grateful to Professor
Xiao-Li Meng for his comments on an earlier version of this paper,
which led to improvements of the paper in several ways. We are
also grateful to two anonymous referees for their valuable
comments that led to a more focused and balanced treatment of the
subjects.

\begin{supplement}[id=suppA]
\stitle{Profiling time course expression of a single virus gene\\}
\slink[doi]{10.1214/09-AOAS258SUPP}
\slink[url]{http://lib.stat.cmu.edu/aoas/258/supplement.pdf}
\sdatatype{.pdf}
\sdescription{This nonhierarchical Bayesian\break method, using also Bernstein
polynomials, allows nontrivial prior probability on the order of
the Bernstein polynomial and is amenable to simulation studies,
which indicate its excellent numerical performance.}
\end{supplement}

\printaddresses


\begin{thebibliography}{99}

\bibitem[\protect\citeauthoryear{}{1994}]{Ayeetel1994}
\textsc{Ayres, M. D., Howard, S. C.,  Kuzio, J., Lopez-Ferber, M.} and
\textsc{Possee, R. D.} (1994). The complete DNA sequence of
\textit{Autographa californica} nuclear polyhedrosis virus.
\textit{Virology} \textbf{202} 586--605.

\bibitem[\protect\citeauthoryear{}{1955}]{Bru1955}
\textsc{Brunk, H. D.} (1955). Maximum likelihood estimates of monotone parameters.
\textit{Ann. Math. Statist.} \textbf{26} 607--616.
\MR{0073894}


\bibitem[\protect\citeauthoryear{}{2007}]{Chaetal2007}
\textsc{Chang, I. S., Chien, L. C., Hsiung, C. A., Wen, C. C.} and \textsc{Wu, Y. J.}
(2007). Shape restricted regression with random Bernstein polynomials.
In \textit{Complex Datasets and Inverse Problems} (R.~Liu, W. Strawderman and C. H. Zhang, eds.).
\textit{IMS Lecture Notes---Monograph Series} \textbf{54} 187--202. Inst. Math. Statist.,
Beachwood, OH.
\MR{2459189}

\bibitem[\protect\citeauthoryear{}{2008}]{Chaetal2008}
\textsc{Chang, I. S., Chien, L. C., Gupta, P. K.,  Wen, C. C., Wu, Y. J.} and \textsc{Hsiung, C. A.} (2008). Supplement to
``Profiling time course expression of virus genes---an
illustration of Bayesian inference under shape restrictions.''


\bibitem[\protect\citeauthoryear{}{2005}]{Chaetal2005}
\textsc{Chang, I. S., Hsiung, C. A., Wu, Y. J.} and \textsc{Yang, C. C.} (2005). Bayesian survival analysis using Bernstein
    polynomials. \textit{Scand. J. Statist.} \textbf{32} 447--466.
\MR{2204629}


\bibitem[\protect\citeauthoryear{}{2006}]{DetNeuPil2006}
\textsc{Dette, H., Neumeyer, N.} and \textsc{Pilz, K. F.} (2006). A simple nonparametric estimator of a strictly monotone regression
function. \textit{Bernoulli} \textbf{12} 469--490.
\MR{2232727}


\bibitem[\protect\citeauthoryear{}{2005}]{Dun2005}
\textsc{Dunson, D. B.} (2005). Bayesian semiparametric isotonic
regression for count data. \textit{J. Amer. Statist. Assoc.}  \textbf{100} 618--627.
\MR{2160564}

\bibitem[\protect\citeauthoryear{}{2005}]{Dupetal2005}
\textsc{Duplessis, M., Russell, W. M., Romero, D. A.} and \textsc{Moineau, S.} (2005).
Global gene expression analysis of two \textit{Streptococcus thermophilus}
bacteriophages using DNA microarray. \textit{Virology}  \textbf{340} 192--208.


\bibitem[\protect\citeauthoryear{}{2001}]{FriMil2001}
\textsc{Friesen, P. D.} and \textsc{Miller, L. K.} (2001). Insect viruses.
In \textit{Fields' Virology}, 4th ed. (D. M. Knipe, P. M. Howley, D. E. Griffin, M. A. Martin,  R. A. Lamb,
B. Roizman and S. E. Straus, eds.) 608--609.  Lippincott Williams \&\ Wilkins, Philadelphia.

\bibitem[\protect\citeauthoryear{}{2004}]{Geletal2004}
\textsc{Gelman, A., Carlin, J. B., Stern, H. S.} and \textsc{Rubin, D. B.} (2004).
\textit{Bayesian Data Analysis}, 2nd ed. Chapman \&\ Hall/CRC, Boca Raton.
\MR{2027492}

\bibitem[\protect\citeauthoryear{}{2003}]{Gel2003}
\textsc{Gelman, A.} (2003).
    A Bayesian formulation of exploratory data analysis and goodness-of-fit
    testing.
    \textit{Int. Statist. Rev.} \textbf{71} 369--382.

\bibitem[\protect\citeauthoryear{}{1996}]{GelMenSte1996}
\textsc{Gelman, A., Meng, X. L.} and \textsc{Stern, H. S.} (1996).
    Posterior predictive assessment of model fitness via realized discrepancies (with discussion).
    \textit{Statist. Sinica} \textbf{6} 733--807.
\MR{1422404}

\bibitem[\protect\citeauthoryear{}{1992}]{GelRub1992}
\textsc{Gelman, A.} and \textsc{Rubin, D. B.} (1992).
    Inference from iterative simulation using multiple sequences
    (with discussion).
    \textit{Statist. Sci.} \textbf{7} 457--511.


\bibitem[\protect\citeauthoryear{}{2003}]{Gij2003}
\textsc{Gijbels, I.} (2003). Monotone regression. Discussion Paper
0334, Institute de Statistique, Universit\'e Catholique de
Louvain. Available at \url{http://www.stat.ucl.ac.be}.

\bibitem[\protect\citeauthoryear{}{2002}]{HalHec2002}
\textsc{Hall, P.} and \textsc{Heckman, N. E.} (2002).
Estimating and depicting the structure of a distribution of random
funcions. \textit{Biometrika} \textbf{89} 145--158.
\MR{1888371}

\bibitem[\protect\citeauthoryear{}{2000}]{HecZam2000}
\textsc{Heckman, N. E.} and \textsc{Zamar, R. H.} (2000).
   Comparing the shapes of regression functions. \textit{Biometrika} \textbf{87} 135--144.
\MR{1766834}

\bibitem[\protect\citeauthoryear{}{1954}]{Hil1954}
\textsc{Hildreth, C.} (1954).
Point estimate of ordinates of concave functions.
\textit{J. Amer. Statist. Assoc.}  \textbf{49} 598--619.
\MR{0065093}

\bibitem[\protect\citeauthoryear{}{1995}]{HilKuzFau1995}
\textsc{Hill, J. E., Kuzio, J.} and \textsc{Faulkner, P.} (1995).
Identification and characterization of the v-cath gene of the baculovirus, CfMNPV.
\textit{Biochimica et Biophysica Acta} \textbf{1264} 275--278.

\bibitem[\protect\citeauthoryear{}{2004}]{Iwaetal2004}
\textsc{Iwanaga, M., Takaya, K., Katsuma, S., Ote, M., Tanaka, S.,
Kamita, S. G., Kang, W. K., Shimada, T.} and \textsc{Kobayashi, M.} (2004).
Expression profiling of baculovirus genes in permissive and
nonpermissive cell lines. \textit{Biochemical and Biophysical
Research Communications} \textbf{323} 599--614.


\bibitem[\protect\citeauthoryear{}{2006}]{Jiaetal2006}
\textsc{Jiang, S. S., Chang, I. S., Huang, L. W., Chen, P. C., Wen, C.
C., Liu, S. C., Chien, L.~C., Lin, C. Y., Hsiung, C. A.} and
\textsc{Juang, J. L.} (2006). Temporal transcription program of
recombinant \textit{Autographa californica} multiple nucleopolyhedrosis
virus. \textit{Journal of Virology} \textbf{80} 8989--8999.

\bibitem[\protect\citeauthoryear{}{1995}]{KasRaf1995}
\textsc{Kass, R. E.} and \textsc{Raftery, A. E.} (1995).
Bayes factors.  \textit{J. Amer. Statist. Assoc.} \textbf{90} 773--795.

\bibitem[\protect\citeauthoryear{}{1996}]{KneMoretal1996}
\textsc{Knebel-Morsdorf, D., Flipsen, J. T., Roncarati, R., Jahnel, F., Kleefsman, A. W.} and \textsc{Vlak, J. M.} (1996).
   Baculovirus infection of \textit{Spodoptera exigua} larvae:
   \textit{lacZ} expression driven by promoters of early genes pe38 and
   me53 in larval tissue. \textit{Journal of General Virology}
   \textbf{77} 815--824.

\bibitem[\protect\citeauthoryear{}{2003}]{Lagetal2003}
\textsc{Lagreid, A., Hvidsten, T. R., Midelfart, H., Komorowski, J.}
and \textsc{Sandvik, A. K.} (2003). Predicting gene ontology biological
process from temporal gene expression patterns. \textit{Genome
Research} \textbf{13} 965--979.

\bibitem[\protect\citeauthoryear{}{1995}]{LavMoc1995}
\textsc{Lavine, M.} and \textsc{Mockus, A.} (1995).
   A nonparametric Bayes method for isotonic regression.
     \textit{J.~Statist. Plann. Inference} \textbf{46} 235--248.

\bibitem[\protect\citeauthoryear{}{1999}]{LavSch1999}
\textsc{Lavine, M.} and \textsc{Schervish, M. J.} (1999).
    Bayes factors: What they are and what they are not.
     \textit{Amer. Statist.} \textbf{53} 119--122.
\MR{1707756}


\bibitem[\protect\citeauthoryear{}{2007}]{Majetal2007}
\textsc{Majtan, T., Halgasova, N., Bukovska, G.} and \textsc{Timko, J.}
(2007). Transcriptional profiling of bacteriophage BFK20:
Coexpression interrogated by ``guilt-by-association'' algorithm.
\textit{Virology} \textbf{359} 55--65.

\bibitem[\protect\citeauthoryear{}{2003}]{Miletal2003}
\textsc{Milks, M. L., Washburn, J. O., Willis, L. G., Volkman, L. E.} and \textsc{Theilmann, D. A.} (2003).
 Deletion of \textit{pe38} attenuates AcMNPV genome replication, budded virus production, and virulence in \textit{Heliothis virescens}.
 \textit{Virology} \textbf{310} 224--234.

\bibitem[\protect\citeauthoryear{}{2008}]{MisChaDas2008}
\textsc{Mishra, G., Chadha, P.} and \textsc{Das, R. H.} (2008).
Serine/threonine kinase (pk-1) is a component of \textit{Autographa
californica} multiple nucleopolyhedrovirus (\textit{Ac}MNPV) very
late gene transcription complex and it phosphorylates a 102 kDa
polypeptide of the complex. \textit{Virus Research} \textbf{137}
147--149.

\bibitem[\protect\citeauthoryear{}{1999}]{Pet1999}
\textsc{Petrone, S.} (1999). Random Bernstein polynomials.
\textit{Scand. J. Statist.} \textbf{26} 373--393.
\MR{1712051}


\bibitem[\protect\citeauthoryear{}{1994}]{ReiGua1994}
\textsc{Reilly, L. M.} and \textsc{Guarino, L. A.} (1994).
The \textit{pk-1} gene of \textit{Autographa californica}
multinucleocapsid nuclear polyhedrosis virus encodes a protein
kinase. \textit{Journal of General Virology} \textbf{75} 2999--3006.

\bibitem[\protect\citeauthoryear{}{2007}]{Smi2007}
\textsc{Smith, I.} (2007). Misleading messengers? Interpreting
baculovirus transcriptional array profiles. \textit{Journal of
Virology} \textbf{81} 7819--7821.

\bibitem[\protect\citeauthoryear{}{2006}]{Munetal2006}
\textsc{van Munster, M., Willis, L. G., Elias, M., Erlandson, M.
A., Brousseau, R., Theilmann, D. A.} and \textsc{Masson, L.} (2006).
Analysis of the temporal expression of \textit{Trichoplusia ni}
single nucleopolyhedrovirus genes following transfection of
BT1-Tn-5B1-4 cells. \textit{Virology} \textbf{354} 154--166.


\bibitem[\protect\citeauthoryear{}{2002}]{Yanetal2002}
\textsc{Yang, W. C., Devi-Rao, G. V., Ghazal, P., Wagner, E. K.} and \textsc{Triezenberg, S. J.} (2002).
General and specific alterations in programming of global viral
gene expression during infection by VP16 activation-deficient
mutants of herpes simplex virus type 1. \textit{Journal of Virology}
\textbf{76} 12758--12774.

\end{thebibliography}
\end{document}